\documentclass[twocolumn]{IETEJR}
\usepackage{array}
\usepackage{amsmath}
\usepackage{cite}
\usepackage{graphicx}
\usepackage[font=small]{caption}
\usepackage[font=small]{subcaption}
\usepackage{algpseudocode}
\usepackage{cuted}
\usepackage{cleveref}
\usepackage{lipsum}
\usepackage{stfloats}
\usepackage{mathcomp}
\usepackage{geometry}
\usepackage{textcomp}
\usepackage{float}
\usepackage{gensymb}
\usepackage{csquotes}
\usepackage{graphics}
\usepackage{xcolor}
\usepackage{epstopdf}
\usepackage{amsmath}
\usepackage{eqnarray}
\usepackage{multirow}
\usepackage{amssymb}
\usepackage{url}
\usepackage[T1]{fontenc}
\usepackage{tfrupee}
\let\orupee\rupee
\def\rupee{\ifmmode\text{\orupee}\else\orupee\fi}
\usepackage{enumitem}
\usepackage{pifont}
\usepackage{url}
\begin{document}
\title{Disturbance Observer Based Frequency \& Voltage Regulation for RES Integrated Uncertain Power Systems} 
\author{ Himanshu Grover, Ashu Verma and T.S.Bhatti\\
\vspace*{0.05em}\\\small Himanshu Grover is with Department of Energy Science and Engineering, Indian Institute of Technology Delhi\\\small New Delhi, 110016 INDIA 
(e-mail:himanshu.grover@ces.iitd.ac.in). \\

\small Ashu Verma, is with Department of Energy Science and Engineering, Indian Institute of Technology Delhi.\\\small New Delhi, 110016 INDIA (\small e-mail: averma@ces.iitd.ac.in). \\
\small T.S.Bhatti is with Department of Energy Science and Engineering, Indian Institute of Technology Delhi\\\small New Delhi, 110016 INDIA (\small e-mail: tsb@ces.iitd.ac.in). \\
    }
\twocolumn[
\begin{@twocolumnfalse}
    \maketitle
    \begin{abstract}
    This paper proposes a disturbance-observer-based control (DOBC) scheme for frequency and voltage regulation for a renewable energy sources (RES) integrated power systems. The proposed approach acts a feed-forward control which improves the dynamic performance of the conventional proportional-integral-derivative (PID) controller. Robustness of the proposed control scheme has been validated through simulations under worst-case and stochastic uncertainties to mitigate real-time variability in RES output and load. The performance of the proposed control technique is compared to well established technique in presence of communication delay and white noise. 
    \end{abstract}
\vspace{1cm}
\end{@twocolumnfalse}
]
\vspace{-0.8cm}
\section{INTRODUCTION}
\indent Thermal power plants constitute a major proportion of generation sources in power systems around the world. These consist of large synchronous generators which regulate the system frequency and voltage due to their high inertia and damping properties \cite{1}. The primary energy source of these generators are fossil fuels which contribute significantly to global carbon emissions and continuous depletion  of fossil fuel reserves, which cannot be replenished in a human time frame \cite{2}. This necessitates adoption of alternate clean generation sources to limit adverse effects on global climate as well as to efficiently regulate power system operation.\\
\indent Recent advancements in generation technologies based on RES such as solar photovoltaic (PV), wind turbine generators and small hydro generation systems have facilitated large-scale integration of RES in modern power systems. RES generation systems are interfaced with power electronic based converters, thereby resulting in large-scale penetration of inverter-interfaced generation sources, leading to reduction in overall inertia and damping of conventional power system. It is estimated that high penetration of RES may result in reduction of nearly 70\% of power inertia during the period of 2014 and 2034 \cite{3}. Moreover, since RES are inherently intermittent and uncertain in nature, their widespread integration causes new challenging tasks for the system controller in terms of operation and reliability. Among these, a major challenge is to ensure frequency and voltage regulation within the specified operational limit. As reduction in the system inertia makes the power system prone to disturbance and outages, it is increasingly susceptible to instability and higher rate of change of frequency \cite{1}.\\     
\indent Large power fluctuations in RES outputs can be minimised using energy storage system (ESS) which minimize active power mismatch and network frequency deviations \cite{4}. Several research works have been carried out in the field of ESS integration with RES. However, installation of high capacity ESS leads to high investment costs and hence, it is not economically viable. This envisages the need to develop a fast dynamic and robust control mechanism for conventional thermal power plants, in order to suitably mitigate the variability and uncertainty of RES/load without using ESS. Conventional power system utilizes integral controller for regulation but it limits the dynamic performance during any large disturbance in the system \cite{5}. PID controller has gained much focus on frequency and voltage regulation. However, the difficulty arises in tuning parameters of PID controllers.\\
\indent In view of the above concerns, a nonlinear-threshold accepting meta-heuristic algorithm is developed in \cite{6} to calculate the PID controller gains for load frequency control (LFC) and automatic voltage regulator (AVR) of synchronous generator. A tuning method for LFC PID controller is proposed in \cite{5} using two-degree-of freedom internal model control (IMC). An application of bacteria foraging optimization algorithm (BFOA) is presented for tuning the PID controller parameters in \cite {7,8}. Various research works have been carried out in tuning the PID controller parameters for LFC/AGC and AVR application in \cite{9,10,11,12}. However, above research did not focus on the impact and challenges in integration of solar PV. \\  
\indent A cascade-$I\lambda D \mu N$ controller has been designed for AGC in \cite{13}. The authors examined the impact of incorporating RES such as solar-thermal, wind etc. to a 2-area multi-source thermal power system. The robustness of the proposed controller was evaluated using sensitivity study considering $\pm 20\%$ deviation in system parameters and load perturbation. However, the authors have not discussed the criteria for selection of the PV output/load deviation as well as the parameter to evaluate robustness.\\
\indent In recent years, the integration of DOBC has been observed to have a great potential in improving dynamic performance in different areas \cite{14,15,16,17}. Authors in \cite {17} developed DOBC based frequency regulation scheme for low-inertia microgrid. The results concludes that the DOBC controller accurately estimates the the mismatch in the system and provides satisfactory results without implementing much computational burden. The DOBC controller can be integrated to existing controller and acts as a feed-forward control strategy, thus improving the dynamic performance of the system. In addition to the above benefits, DOBC does not require any additional sensor to sense the disturbance thus it is economical viable. \\    
\indent To the best of the authors’ knowledge, existing research does not focuses on performance evaluation of large thermal-solar PV system under system uncertainty. Keeping in view the issues arising in large-scale RES integration in the conventional power system, a fast dynamic frequency and voltage regulation scheme has been proposed in this work. Accordingly, this paper focuses on DOBC-based frequency and voltage regulation scheme to mitigate real-time uncertainties arising due to variability in solar PV output and load. The contribution of the paper are as follows.
\begin{enumerate}
    \item Implementation of DOBC as an auxiliary control to existing PID controller for frequency and voltage regulation, which improves the system performance without rendering much computational burden.
    \item The proposed control scheme has been simulated for different levels of PV and load uncertainties quantified as worst-case and stochastic variations, to validate operational robustness during real-time implementation.
    \item Verification of proposed scheme considering practical scenarios such presence of white noise and communication delay.
    \end{enumerate}
\indent The  remainder of the paper  is  organized  as  follows. The system  description is provided in Section II. Section III provides the detailed modelling and designing of DOBC for its application in frequency-voltage regulation. Section IV provides simulation results and discussion. Finally conclusion of research work is drawn in Section V.
\section{SYSTEM DESCRIPTION}
\begin{figure}\centering
	\includegraphics[width=0.95\linewidth]{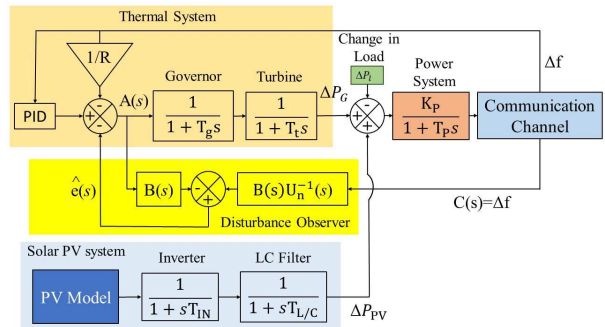}
	\caption{Block diagram of LFC for a single area power system.}
	\label{LFC}
\end{figure}
\begin{figure}[h]\centering
	\includegraphics[width=\linewidth]{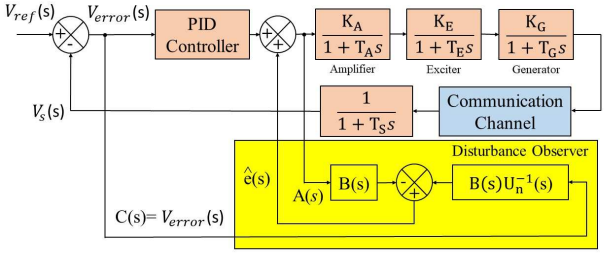}
	\caption{Block diagram of AVR for a single area power system.}
\vspace{-0.6cm}
	\label{AVR}
\end{figure}

The block diagram of load frequency Control (LFC) and automatic voltage regulation (AVR) for a single area power system is shown in Fig. \ref{LFC} and \ref{AVR} respectively. 
The LFC system contains the non-reheat thermal power system integrated with solar PV generator, thermal power system handles the intermittent nature of solar PV and loads. The synchronous generator of thermal power system regulates the system frequency and voltage within the prescribed limits. $\Delta P_l$, $\Delta P_G$  and $\Delta f$ refer to changes in load power, generator output and frequency, respectively for the thermal power system. $T_{g}$ is speed governor time constant, $T_{t}$ is turbine time constant and R is the regulation parameter of primary controller \cite{18}. Solar PV system consists PV model connected through inverter and L-C filter. The inverter is a power electronic device which converts the DC power generated from the solar PV to AC power. The inverter operates in grid following mode and follows the frequency and voltage generated by the synchronous generator.\\
\indent In this paper an equivalent solar PV model for a large distributed solar PV farm is considered, however a first order element of inverter and LC filter in s-domain are considered. The time constant of inverter and LC filter represents the time which a inverter takes to change the power \cite{1}.
$ T_{IN} $, $T_{L/C}$ and $\Delta P_{PV}$ are inverter time constant, inverter L/C filter time constant and change in the PV output respectively for the PV generation system. Similarly, Fig. \ref{AVR} illustrates the block diagram of automatic voltage regulation (AVR) for a single area power system where $V_{ref}(s)$, $V_s(s)$ and $V_{error}(s)$ refer to reference voltage, sensor output voltage (actual system voltage) and the error in voltage respectively. $K_A$ is gain and $T_A$ is time constant of amplifier, $K_E$ is gain and $T_E$ is time constant of exciter. $K_G$ is gain and $T_G$ is time constant of generator and $T_S$ is time constant of sensor \cite{9}.
\\ \indent It is a assumed that for a large power system, the active power mismatch in the system primarily affect the system frequency, whereas reactive power mismatch in the affects the bus voltage magnitudes \cite{18}. The frequency deviation in power system is obtained by power system dynamics model and its parameters are calculated by the following \cite{19};  
\begin{gather}
 K_p \triangleq\frac{1}{D} \;\;\;\;\;  Hz/pu MW\\
 T_p\triangleq \frac{2H}{f^\circ}D\;\;\;\; s\;\;\;\;\;\;\;\;\;\;\;\;\;\;
\end{gather}
where, D is damping factor in Hz/pu MW, $f^\circ$ is nominal frequency in Hz and  H is inertia constant in seconds, which is the ability of a system to resist change in frequency during any contingency. 
Any mismatch in the generated power and demand power can lead to frequency deviation in the system. The frequency deviation majorly depends on system inertia (H) and damping factor (D). The system frequency dynamics can be governed by the first-order swing equation \cite{20}:
\begin{gather}
\frac{d\Delta f}{dt}=\frac{1}{2H}[\Delta P_{DG}(t)-\Delta P_L(t)-D.\Delta f(t)]
\end{gather}
The power supplied to the load is the sum of power generated by thermal power system and solar PV generator. However, the output power of solar PV generator is intermittent and fluctuating in nature due to variable climatic conditions. Also, the load demand changes all the time which leads to an imbalance in generated power and demand power. If $P_L$ is power supplied to load including losses, $P_{G}$ is power generated by thermal system and $P_{PV}$ is power generated by solar PV, the power balance equation for the system during islanding condition is expressed as:
\begin{gather}
P_L=P_{G}+P_{PV} 
\end{gather}
Active power balance and frequency regulation of a system is referred as load frequency contol (LFC) \cite{21}. LFC tends to balance active power output between thermal power system, solar PV and load during islanding condition. Any power imbalance in load or source (due to change in solar irradiance or fuel input for thermal power system) leads to frequency deviation which can be measured using a first order transfer function of power system given by \cite{19}:
\begin{gather}
G_p(s)=\frac{K_p}{1+sT_p}\;\;\;\;\;\;\;\;\;\;\;\;\;\;\;
\end{gather}
where $K_p$ is power system gain and $T_p$ is power system time constant.\\
\begin{figure*}[b]
\hrule
\vspace{0.5cm}
\begin{equation}
\frac{\Delta V_S(s)}{\Delta V_{ref}(s)} = \frac{(s^2K_d+sK_p+K_i)(K_aK_eK_g)(1+sT_s)}{s(1+sT_a)(1+sT_e)(1+sT_g)(1+sT_s)+(K_aK_eK_gK_s)(s^2K_d+sK_p+K_i}  \label{TF}
\end{equation}
\end{figure*}
\indent  The AVR system as shown in Fig. \ref{AVR} is faster and less complex as compared to LFC control loop. The AVR for thermal power system contains controller, amplifier, exciter, sensor and generator. The sensor sense the terminal voltage of generator and sends the data to controller using a communication channel. The controller regulates the generator voltage based on reference signal by controlling the field winding of generator. The transfer function of overall AVR system for a single area power system is represented as \eqref{TF}\cite{9}.
\vspace{-0.5cm}
\subsection{Quantification of uncertainty occurrences}
\indent Increasing penetration of solar PV introduces generation uncertainties in the power system. This can be attributed to the inherent variability associated with the solar resource. Similarly, uncertainties are observed in the electrical load arising out of the increased proportion of fast-switching consumer loads. In order to evaluate the impact of these generation and load uncertainties on power system performance, this work employs two most common uncertainty quantification techniques available in literature, which are \textit{(a) worst-case} and \textit{(b) stochastic} uncertainties.
\subsubsection{Worst-case uncertainty}
\indent The worst-case uncertainties are evaluated through a robust optimization formulation, based on a $max-min$ optimization model. The $min$ operator minimizes the function value for maximum (worst-case) occurrence of uncertainty obtained by the $max$ operator. The uncertain variable $(u)$ is defined as a deterministic set $(U)$ bound by lower and upper limits, referred to as uncertainty budgets. Essentially, the worst-case value of $u$ refers to its value within the pre-defined uncertainty set, for which the dual of the inner $min$ function achieves a maxima. In context of LFC, the worst-cases of load and PV output correspond to the values from within their respective uncertainty sets which result in the maximum value of $\Delta$f. Though uncertainty sets maybe geometrically described using different types of uncertainty sets, this work assumes the uncertainty sets to be polyhedral uncertainties, which ensure a suitable trade-off between solution robustness and computational efficiency. Mathematical representation of the polyhedral uncertainty set is given as follows \cite{1}.
\begin{eqnarray}
\begin{aligned}
U=\{u^t \in {\rm I\!R}^{n} :
u^{t,{min}}\le u^t \le u^{t,max}, \forall \ t \\
\mu^l \le \frac{\sum\limits_{t \in T} u^t}{\sum\limits_{t \in T} u^{f,t}} \le \mu^u \}
\end{aligned}
\end{eqnarray}
\indent Here, $u^{t,min/max}$ and $u^{t,f}$ refer to the minimum/maximum and forecast value of $u$ at time $t$, respectively. $\mu^{l/u}$ refers to the lower/upper limits of the uncertainty set $U$.
\subsubsection{Stochastic uncertainty}
\indent Though uncertainty quantification as worst-case occurrences ensures robust system operation, it is noteworthy that worst-cases are rarely encountered in practice. Another method for quantification of uncertainties is based upon probabilistic analysis of uncertainty occurrences. In case sufficient information about probability distribution of the uncertain variables is available, stochastic uncertainty scenarios maybe developed to represent the occurrence of uncertainties. It has been well established in literature that stochastic variation in solar PV output follows the beta distribution function, while electrical load uncertainty has been observed to follow normal distribution. Mathematically, this may be expressed as follows \cite{22}.
\begin{eqnarray}
f_{pv}(y)=y^{\alpha-1}(1-y)^{\beta-1}N_\beta \label{c4pvun}\\
f_{l}(y)=\frac{1}{\sqrt{2\pi(\sigma)^2}}exp(-\frac{(x-\mu)^2}{2\sigma_l^2}) \label{c4loadun}
\end{eqnarray}
\indent where, $\alpha$ and $\beta$ are shape parameters of the beta distribution, $N_\beta$ is a normalization factor, and $y$ is the occurrence of $pv$; $\mu$ and $\sigma$ are the mean and standard deviation of the load forecast, respectively.\\
\indent Evaluation of system performance under stochastic uncertainties requires generation of a sufficiently high number of uncertainty scenarios through Monte Carlo sampling technique. However, evaluation of a large number of samples, especially at the scale of the power system, is a computationally challenging task. In order to reduce the computational burden, scenario reduction techniques are employed which reduce the number of samples to a smaller, yet representative set of uncertainty. In this paper, the backward reduction technique has been employed which has been elaborately discussed in \cite{23}.
\section{DISTURBANCE OBSERVER BASED CONTROL STRATEGY}
\subsection{DOBC Methodology}
\begin{figure}[h]\centering
	\includegraphics[width=\linewidth]{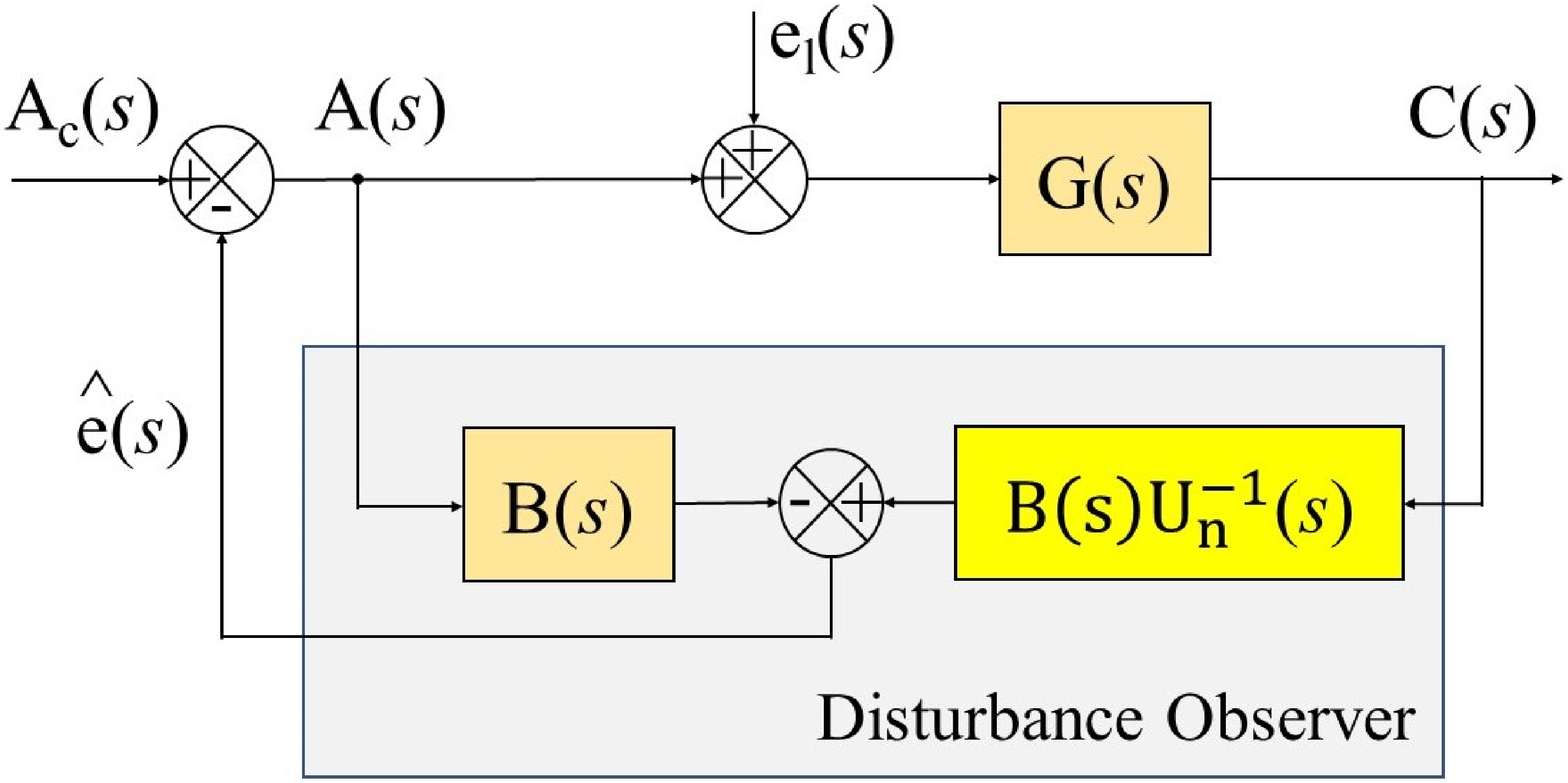}
	\caption{Frequency domain disturbance observer architecture.}
	\label{DOBC}
\end{figure}

In this paper a single-input single-output (SISO) linear disturbance estimator in frequency domain form is considered as shown in Fig. \ref{DOBC}.
The DOBC architecture in frequency domain is explained as \cite{24}:
\begin{equation}
   C(s)= U(s)[A(s)+e_l(s)]
\end{equation}
where A(s) is the control input, $e_l(s)$ is the lumped disturbance value, U(s) is the plant transfer function, C(s) is the controlled output. 
DOBC estimates both internal and external disturbances due to presence of model uncertainties. Therefore, DOBC architecture is modified which consists of lumped disturbance. The estimation of lumped disturbance is evaluated as:
\begin{equation}
e_l(s)=U_n^{-1}(s)U(s)e(s)+[U_n^{-1}(s)U(s)-1]A(s)
\end{equation}
Where e(s) is system disturbance, $U_n^{-1}$ is the inverse of nominal plant transfer function. In the absence of any disturbance in the system the plant transfer function U(s) is same as inverse of nominal plant transfer function $U_n^{-1}$. The estimated equivalent disturbance is evaluated as:
\begin{equation}
\hat{e}(s)=B(s)U_n^{-1}(s)C(s)-B(s)A(s)\\
\end{equation}
\begin{equation}
          =B(s)U_n^{          -1}(s)U_n(s)(A(s)+e_l(s))-B(s)A(s)\\
\end{equation}
\begin{equation}
          =B(s)e_l(s)
\end{equation}
where $\hat e(s)$ is the estimated value of equivalent disturbance. Error $E_d(s)$ in estimated lumped disturbance is evaluated as difference in estimated equivalent disturbance $\hat e(s)$ and lumped disturbance $e_l(s)$.  
\begin{equation}
E_d(s)=\hat{e}(s)-e_l(s)=[B(s)-1]e_l(s)\\
\end{equation}
\begin{equation}
=[B(s)-1]e_l(s)
\end{equation}

Design of filter B(s) as a low-pass filter (LPF) results in the value of lumped disturbance estimation error tending to zero as time tends to infinity.

\subsection{Designing and implementation of Linear Disturbance Estimator for LFC}
The designing and implementation of linear disturbance estimator for LFC is provided in this section. As per the earlier assumptions, the disturbance (e(s)) in the system affects the control input (A(s)). However, for LFC application the disturbance(e(s)) does not directly affects the control instructions \cite{17}.
The disturbance(e(s)) for LFC is the resultant of active power mismatch between solar PV and load, 
\begin{equation}
e(s)={\Delta P_{PV}(s)- \Delta P_L(s)}
\end{equation}
The ‘+’ ve sign before $ P_{PV} $ is taken as it generates power and ‘-’ve sign before $ \Delta P_L $ as it consumes power.

Apart from the identification of lumped disturbance $e_l(s)$ in the system. The design process of implementing DOBC for
LFC includes two more procedures which are as follows:
(1) identification of overall plant transfer function (U(s)); 
(2) filter designing (B(s))

\begin{figure}[h]\centering
	\includegraphics[width=\linewidth]{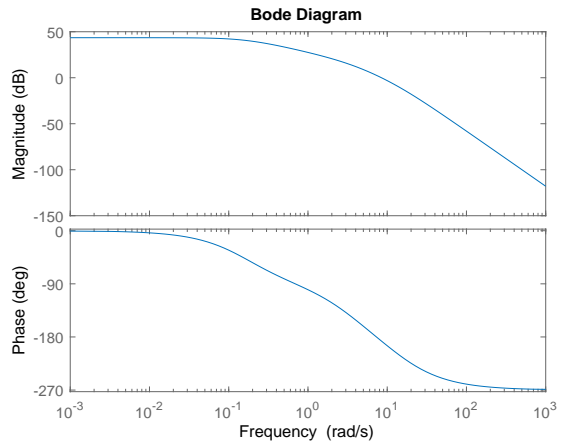}
	\caption{Bode diagram of plant transfer function.}
	\label{Bode1}
\end{figure}

\begin{figure}[h]\centering
	\includegraphics[width=\linewidth]{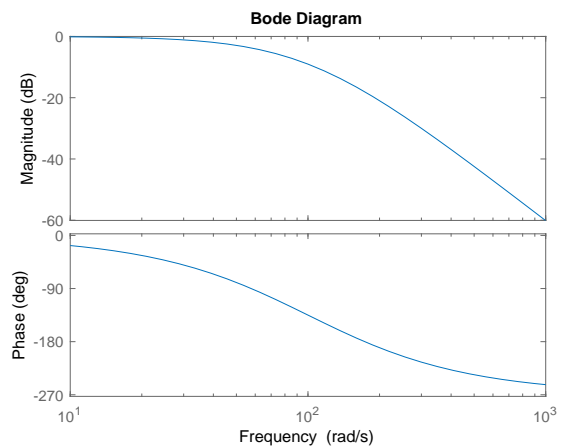}
	\caption{Bode diagram of filter transfer function.}
	\label{Bode2}
\end{figure}

(1) Identification of overall plant transfer function: The overall plant transfer function is given as;
\begin{equation}
    U(s)=U_P(s).U_{T}(s).U_{G}(s)
\end{equation}
\begin{equation}
=\Bigg[\frac{K_P}{1+sT_P}\Bigg]\Bigg[\frac{K_t}{1+sT_t}\Bigg]\Bigg[\frac{K_g}{1+sT_g}\Bigg]
\end{equation}
(2) filter designing (B(s)): 
Performance of DOBC majorily depends upon the filter design. Filter must be designed as LPF such that it must eliminate the high frequency noise signal generated via sensor \cite{17}. The filter must be designed keep in concern that the order of the filter must be equal to or greater than the difference in order of numerator and denominator of plant transfer function. The order ensures that control structure is realizable i.e. $B(s)U^{-1}_n(s)$ must be proper.  
In addition to selection of order, the selection of parameter for B(s) is such that in low-frequency domain $B(s) \cong 1$. It guarantees the estimation of lumped disturbance is nearly equals to actual lumped disturbance $e_l(s)$.  
Keeping in concern the above conditions for designing the filter, the order of plant for LFC should not be less than third order with a steady state gain of 1. The third order transfer function of filter is selected in this case given by:
\begin{equation}
  B(s)= \frac{1}{(\lambda s+1)^3}
\end{equation}
The filter accuracy in estimating disturbance depends upon the selection of filter parameter $\lambda$.
The bode diagram of overall plant function and filter is represented in Fig. \ref{Bode1} \& \ref{Bode2} respectively. It states that the frequencies higher than $10^{-1} rad/s$ are attenuated. Since the phase lag present in control loops deteriorate the performance of system by making it slower, therefore, the system must have zero phase lag up to $10^{-1} rad/s$.
The values of gain, phase and cut-off frequency at different values of lambda ($\lambda$) are listed in Table \ref{Filter}. Keeping in view of the need to maintain unity gain and nearly zero phase gap at frequency $10^{-1} rad/s$, the value of lambda ($\lambda$) is chosen as 0.01.
\begin{table*}[h]
	\centering\small\caption{Filter performance at different values of $\lambda$.}\label{Filter}
	\begin{tabular}{c c c c}
		\hline 
		\hline 		
		\rule{0pt}{2ex} \textbf{Lambda} & \textbf{Gain(dB)} & \textbf{Phase(degree)} & \textbf{Cut-off} \\
		\rule{0pt}{2ex} \textbf{$\lambda$} & \textbf{$@10^{-1}rad/sec$} & \textbf{$@10^{-1}rad/sec$} & \textbf{frequency} \\
		\hline
		\rule{0pt}{2ex} 5 & -3.28 & -84.8 & 0.1018 \\
		\rule{0pt}{2ex} 4 & -1.85 & -64.1 & 0.1272 \\
		\rule{0pt}{2ex} 3 & -1.02 & -47.9 & 0.1696 \\
		\rule{0pt}{2ex} 2 & -0.558 & -35.4 & 0.2544 \\
		\rule{0pt}{2ex} 1 & -0.162 & -19.1 & 0.5088 \\
		\rule{0pt}{2ex} 0.5 & -0.034 & -8.78 & 1.0177 \\
		\rule{0pt}{2ex} 0.2 & $-7.12*10^{-3}$  & -4.02 & 2.5442 \\
		\rule{0pt}{2ex} 0.15 & $-5.21*10^{-3}$  & -3.44 & 3.3923 \\
		\rule{0pt}{2ex} 0.1 & $-3.65*10^{-3}$ & -2.506 & 5.0885 \\
		\rule{0pt}{2ex} 0.05 & $-2.10*10^{-3}$  & -1.575 & 10.1769  \\
		\rule{0pt}{2ex} 0.01 & $-1.31*10^{-3}$ & -0.3102 & 26.91 \\
		\hline
	\end{tabular}
\end{table*}
Similar procedure has been carried out for the identification of overall plant transfer function and filter designing for voltage regulation. It can observed that the designing of DOBC for frequency and voltage regulation does not depends upon the secondary controller. So it can be stated that implementation of DOBC for frequency and voltage regulation is independent of controller and it can be integrated to any existing controller to improve its performance.

\section{SIMULATION RESULTS}
Dynamic performance of the proposed DOBC controller for LFC and AVR under various scenarios of step load change and PV output variation has been elaborately discussed and analyzed in this section.
The proposed DOBC strategy is simulated on MATLAB/Simulink platform. 
The impact of communication delay and presence of white noise is also evaluated. 
Controller performance has been evaluated using performance indices (PI) such as integral of the square of the error (ISE), integral of time multiplied square of the error (ITSE), integral of the time multiplied absolute value of error (ITAE), integral of the absolute value of error (IAE) and maximum overshoot (MO).
\subsection{Performance evaluation of LFC}
\begin{figure}[h]\centering
	\includegraphics[width=\linewidth]{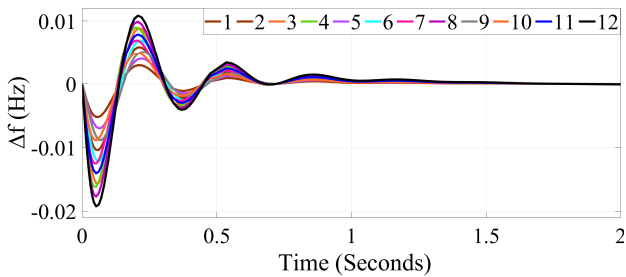}
	\caption{Frequency response for robust uncertainty test conditions.}
	\label{UncertainityBudgets}
\end{figure}

\begin{figure}[h]\centering
	\includegraphics[width=\linewidth]{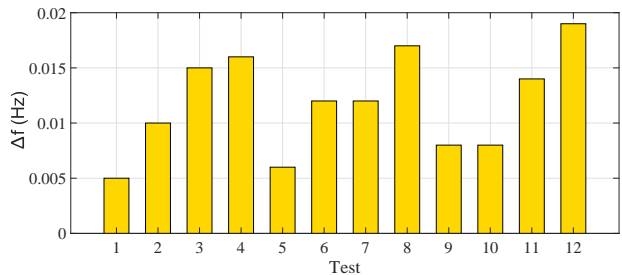}
	\caption{Maximum frequency deviation under different test conditions.}
	\label{DelF}
\end{figure}

\begin{figure}[h]\centering
	\includegraphics[width=\linewidth]{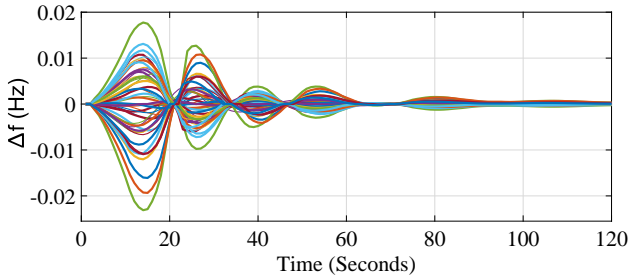}
	\caption{Frequency  response  for  stochastic uncertainty test conditions.}
	\label{StochasticUncertainity}
\end{figure}

\begin{figure}[h]\centering
	\includegraphics[width=\linewidth]{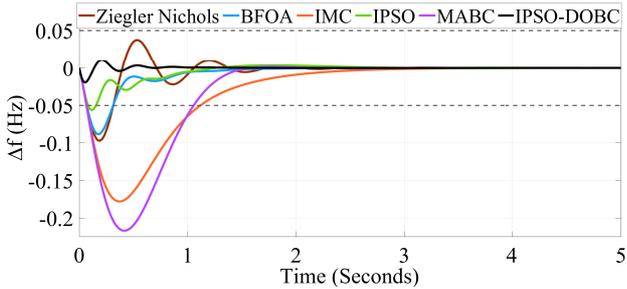}
	\caption{Frequency response for a step change in solar PV and load.}
	\label{FStepchange}
\end{figure}

\begin{figure}\centering
	\includegraphics[width=\linewidth]{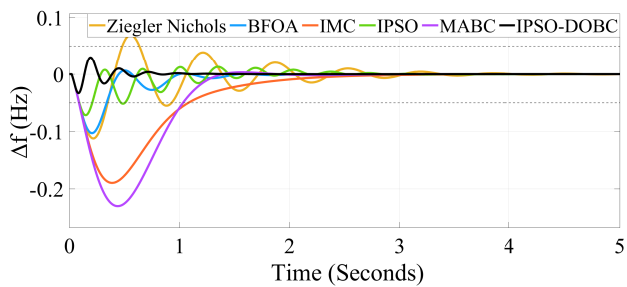}
	\caption{Frequency response under communication delay.}
	\label{Fdelay}
\end{figure}

\begin{figure}\centering
	\includegraphics[width=\linewidth]{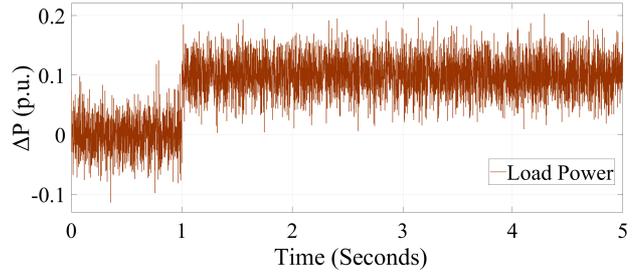}
	\caption{Step load disturbance with white noise.}
	\label{LWhitenoise}
\end{figure}

\begin{figure}\centering
	\includegraphics[width=\linewidth]{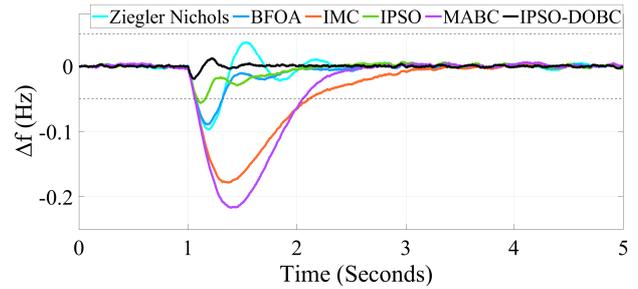}
	\caption{Frequency response with white noise.}
	\label{Fwhitenoise}
\end{figure}

Robustness of the proposed approach is evaluated by identifying worst case scenario. To identify worst-case, uncertainty budgets under different test conditions were considered as listed in Table \ref{budget}, where $\gamma_{PV}^l$ is the lower \& $\gamma_{PV}^u$ is the upper limit of PV deviation in \textit{pu}, $\Delta P_{PV}$ is step change in PV power in \textit{pu} considering base value of thermal system, $\gamma_{L}^l$ is lower \& $\gamma_{L}^u$ is upper limit of load deviation in \textit{pu}, $\Delta P_{L}$ is step change in load power in \textit{pu} considering base value of thermal system. $\Delta f_{max}$ is maximum frequency deviation in Hz. 
Values of $\Delta f_{max}$ under  different  test  conditions  are  listed  in  Table \ref{budget}. The frequency response for different test conditions is depicted in Fig. \ref{UncertainityBudgets}. A comparison of maximum frequency deviation under different test conditions is shown in Fig. \ref{DelF} and found to be maximum for test no. 12, while considering test no. 12 as worst case uncertainty scenario.
Furthermore, the proposed control strategy has been evaluated under occurrence of stochastic uncertainties in PV output and load, considering beta probability distribution function (PDF) for PV uncertainty and normal PDF for load uncertainty. Random uncertainty scenarios were generated using Monte Carlo simulations, and reduced to a small representative set of the uncertainty scenarios.  
Frequency response under stochastic uncertainties is shown in Fig. \ref{StochasticUncertainity}. It has been observed from Figs. \ref{UncertainityBudgets} \& \ref{StochasticUncertainity}, that even in presence of worst-case and stochastic uncertainties, the proposed controller is able to suitably mitigate frequency deviations within prescribed limits. It can observed from the frequency response, that the proposed control strategy regulates the frequency deviation within $5\%$ of its nominal limits. \\
\indent Comparison of dynamic performance in integration of DOBC with existing controller with well established techniques \cite{5,8,10} is shown in this section. The PID controller is integrated with DOBC controller for frequency regulation and the parameters of PID controllers are considered in \cite{10}, which are evaluated using Improved Particle Swarm Optimization (IPSO). 
Fig. \ref{FStepchange} depicts dynamic frequency response of different controllers for step change in PV output and load for the conditions as considered in test no. 12. 
Detailed comparison of the performance of different controllers is presented in Table \ref{step} and PI are found to be minimum for DOBC integrated with PID parameters of IPSO algorithm. As it can be observed in the case of frequency regulation the values of PI in integration of DOBC with PID (parameters from IPSO) is found to be minimum.\\
\indent For a large power system the sources are distributed at different geographical locations and the sensors installed at different sources communicates the information of electrical parameters for safe and reliable operation. For a practical system it can be assumed that the system may face random communication delay in transferring information from one location to another. It is stated that the presence of random delay may degrade the system performance and make the system unstable \cite{17}. Therefore the performance is evaluated by testing the system at random communication delay, Fig. \ref{Fdelay} shows the frequency response for step change for conditions considered in test no. 12 under communication delay of t= 0.02s, and comparison of PI for the proposed technique with existing techniques are compared in Table \ref{delay}. The PI of frequency regulation under communication delay are found to be minimum for proposed technique. So, it can be stated that the proposed technique provides superior performance during the presence of communication delay.\\
\indent The dynamic performance of the proposed technique is evaluated under step change in load having white noise which represents a real-world scenario. Typically, white noise is commonly found in power system due to frequent switching operation of loads \cite{25}. 
A step change in PV and load is considered for the conditions as stated in test no. 12. A 0.1 \textit{pu} step load disturbance with white noise is shown in Fig. \ref{LWhitenoise} and frequency response in the presence of white noise is shown in Fig. \ref{Fwhitenoise}. The PI of different techniques are compared in Table \ref{whitenoise} and it can be concluded that the proposed technique demonstrated better dynamic performance under the presence of white noise is load as compared to existing techniques and PI are found to be minimum for proposed technique. 
\vspace{-0.3cm}
\vspace{-0.3cm}
\begin{table*}[h]
\caption{Uncertainty budgets under different test conditions.}\label{budget}\centering\small
   \begin{tabular}{c c c c c c c c}
        \hline
        \hline
        Test & $\gamma_{PV}^l$  & $\gamma_{PV}^u$ & $\Delta P_{PV}$ & $\gamma_{L}^l$ & $\gamma_{L}^u$ & $\Delta P_{L}$ & $\Delta f_{max} $ \\
        \hline
        \rule{0pt}{2ex}1 & $0.95$ & $1.05$ & $0.0187$ & $0.95$ & $1.05$ & $0.025$ & 0.005 \\
        \rule{0pt}{2ex}2 & $0.90$ & $1.10$ & $0.0375$ & $0.90$ & $1.10$ & $0.050$ & 0.01 \\
        \rule{0pt}{2ex}3 & $0.85$ & $1.15$ & $0.0562$ & $0.85$ & $1.15$ & $0.075$ & 0.015 \\
        \rule{0pt}{2ex}4 & $0.95$ & $1.05$ & $0.0187$ & $0.80$ & $1.20$ & $0.100$ & 0.016 \\
        \rule{0pt}{2ex}5 & $0.90$ & $1.10$ & $0.0375$ & $0.95$ & $1.05$ & $0.025$ & 0.006 \\
       \rule{0pt}{2ex}6 & $0.85$ & $1.15$ & $0.0562$ & $0.90$ & $1.10$ & $0.050$ & 0.012 \\
        \rule{0pt}{2ex}7 & $0.95$ & $1.05$ & $0.0187$ & $0.85$ & $1.15$ & $0.075$ & 0.012 \\
        \rule{0pt}{2ex}8 & $0.90$ & $1.10$ & $0.0375$ & $0.80$ & $1.20$ & $0.100$ & 0.017 \\
        \rule{0pt}{2ex}9 & $0.85$ & $1.15$ & $0.0562$ & $0.95$ & $1.05$ & $0.025$ & 0.008 \\
        \rule{0pt}{2ex}10 & $0.95$ & $1.05$ & $0.0187$ & $0.90$ & $1.10$ & $0.050$ & 0.008 \\
        \rule{0pt}{2ex}11 & $0.90$ & $1.10$ & $0.0375$ & $0.85$ & $1.15$ & $0.075$ & 0.014 \\
        \rule{0pt}{2ex}12 & $0.85$ & $1.15$ & $0.0562$ & $0.80$ & $1.20$ & $0.100$ & 0.019 \\
        \hline
        \hline
    \end{tabular}
\end{table*}
\vspace{-0.25cm}
\vspace{-0.25cm}
\begin{table*}[h]
\caption{Comparison of transient response analysis of the LFC system for a step change in PV and load.}\label{step}\centering
    \begin{tabular}{c c c c c c}
        \hline
        \hline
         & ITAE & IAE & ISE & ITSE & MO \\
        \hline
        \rule{0pt}{2ex}Ziegler Nichols & $0.01956$ & $0.04001$ & $0.00208$ & $0.00055$ & $0.097$ \\
        \rule{0pt}{2ex}BFOA  & $0.01365$ & $0.03343$ & $0.00166$ & $0.00038$ & $0.088$ \\
        \rule{0pt}{2ex}IMC & $0.10850$ & $0.15340$ & $0.01771$ & $0.00904$ & $0.178$ \\
        \rule{0pt}{2ex}IPSO & $0.01656$ & $0.02592$ & $0.00065$ & $0.00018$ & $0.056$ \\
        \rule{0pt}{2ex}MABC & $0.08742$ & $0.15730$ & $0.02479$ & $0.01214$ & $0.216$ \\
       \rule{0pt}{2ex}IPSO-DOBC & $0.001641$ & $0.00427$ & $0.00004$ & $0.000005$ & $0.019$ \\
        \hline
        \hline
    \end{tabular}
\end{table*}
\begin{table*}
\caption{Comparison of transient response analysis of the LFC system for a step change in PV and load under communication delay.}\label{delay}\centering
    \begin{tabular}{c c c c c c}
        \hline
        \hline
         & ITAE  & IAE & ISE & ITSE & MO \\
        \hline
        \rule{0pt}{2ex}Ziegler Nichols  & $0.0759$ & $0.0795$ & $0.0040$ & $0.00210$ & $0.12$ \\
        \rule{0pt}{2ex}BFOA  & $0.0144$ & $0.0346$ & $0.0021$ & $0.00050$ & $0.10$ \\
        \rule{0pt}{2ex}IMC  & $0.1860$ & $0.1533$ & $0.0186$ & $0.00090$ & $0.19$ \\
        \rule{0pt}{2ex}IPSO  & $0.0255$ & $0.0033$ & $0.0011$ & $0.00040$ & $0.07$ \\
        \rule{0pt}{2ex}MABC & $0.0888$ & $0.1582$ & $0.0265$ & $0.01320$ & $0.23$ \\
       \rule{0pt}{2ex}IPSO-DOBC & $0.0027$ & $0.0081$ & $0.0001$ & $0.00002$ & $0.03$ \\
        \hline
        \hline
    \end{tabular}
\end{table*}
\begin{table*}\caption{Comparison of transient response analysis of the LFC system with step change in solar PV and load with white noise.}\label{whitenoise}\centering
    \begin{tabular}{c c c c c c}
        \hline
        \hline
         & ITAE & IAE & ISE & ITSE & MO \\
        \hline
        \rule{0pt}{2ex}Ziegler Nichols & $0.07384$ & $0.0443$ & $0.00205$ & $0.00263$ & $0.097$ \\
        \rule{0pt}{2ex}BFOA & $0.06061$ & $0.0379$ & $0.00168$ & $0.00210$ & $0.089$ \\
        \rule{0pt}{2ex}IMC & $0.26910$ & $0.1578$ & $0.01803$ & $0.018030$ & $0.170$ \\
        \rule{0pt}{2ex}IPSO & $0.05140$ & $0.0289$ & $0.00066$ & $0.00088$ & $0.056$ \\
        \rule{0pt}{2ex}MABC & $0.26380$ & $0.1653$ & $0.02522$ & $0.03760$ & $0.216$ \\
       \rule{0pt}{2ex}IPSO-DOBC & $0.01518$ & $0.0072$ & $0.00004$ & $0.00005$ & $0.020$ \\
        \hline
        \hline
    \end{tabular}
\end{table*}
\vspace{1.0cm}
\subsection{Performance evaluation of AVR}
\begin{figure}{h}\centering
	\includegraphics[width=\linewidth]{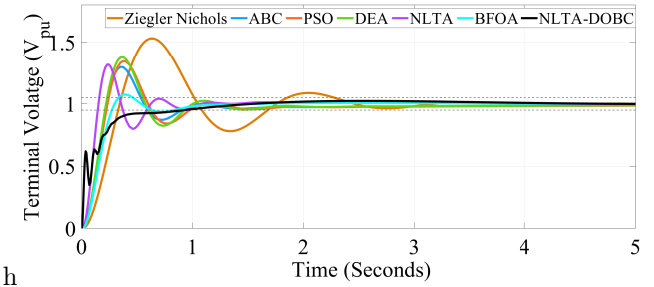}
	\caption{Terminal voltage response.}
	\label{VR}
\end{figure}

\begin{figure}{h}\centering
	\includegraphics[width=\linewidth]{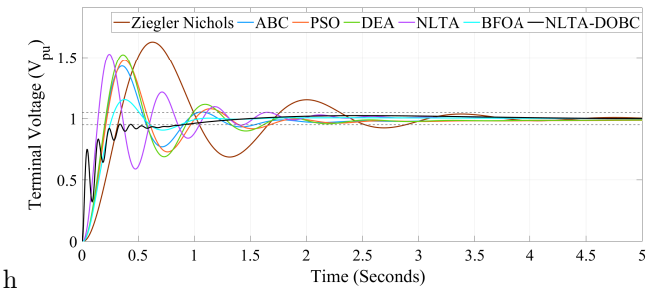}
	\caption{Terminal voltage response under communication delay.}
	\label{Vdelay}
\end{figure}

\begin{figure}{h}\centering
	\includegraphics[width=\linewidth]{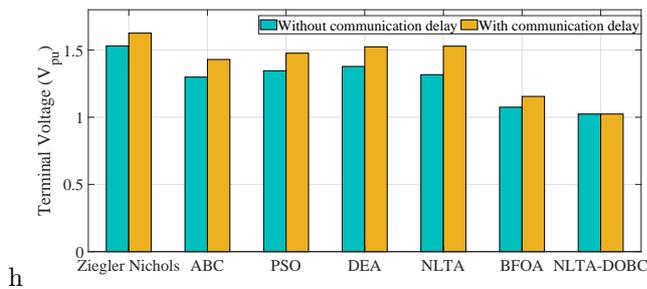}
	\caption{Maximum overshoot with and without communication delay.}
	\label{MOdelay}
\end{figure}

Similarly, two voltage regulation cases are considered in this section to validate the effectiveness of integrating DOBC with conventional PID converter. The parameters of PID values considered in integration with DOBC are considered in \cite{6}. A reference value of 1.0 p.u. is provided at time t=0 seconds and the dynamic performance of the proposed technique is compared with the existing techniques are shown in \cite {6,8,9}. Fig. \ref{VR} shows the terminal voltage response without any communication delay and Fig. \ref{Vdelay} shows terminal voltage response with a communication delay of \textit{t=0.02s}. It can be observed that the proposed DOBC control, when integrated with PID (parameters from NLTA) outperform other well-tuned controllers. Similarly, the maximum overshoot of terminal voltage with and without communication delay is shown in Fig. \ref{MOdelay} and minimum terminal voltage overshoot is found in proposed NLTA-DOBC control.\\ 
\vspace{-0.70 cm}
\section{CONCLUSION}
A fast dynamic frequency and voltage regulation scheme has been developed for a PV integrated single area power system. Disturbance-observer based control has been applied to the conventional PID controller of a synchronous generator. This improves the dynamic performance of the PID controller, which has been validated through comparison with various well-established tuning algorithms. Furthermore, the developed control scheme performs efficiently in RES-penetrated
systems with high volatility of power supply. This has been verified through comprehensive tests under worst-case and stochastic uncertainties in RES output and load. Extensive simulations reveals superior system performance under noisy operating conditions with communication delays.




\section*{APPENDIX}
\subsection{Parameters of thermal power system}
$K_g$ = 1, $T_g$ = 0.08, $K_g$ = 1,$T_t$ = 0.3, $K_A$ = 10, $T_A$ = 0.1, $K_E$ = 1, $T_E$ = 0.4, $K_G$ = 1, $T_G$ = 1, $T_S$ = 0.01, Rated Area Power= 2000 MW, Nominal Operating Load= 1000 MW.
\subsection{Parameters of solar PV system}
$T_IN$ = 0.04, $T_L/C$ = 0.004, Rated Power $(@ 1000 W/m^2)$ = 1000 MW .
\subsection{Parameters of power system}
$K_P$ = 120Hz/pu MW, $T_P$ = 20s, $H$ = 5.0 s , $D$ = 0.00833 pu MW/Hz, $f^\circ$ = 60 Hz.
\begin{table}[h]
\caption{Parameters of PID Controller for LFC}\label{LFC_PID}\centering
    \begin{tabular}{c c c c}
        \hline
        \hline
         & $K_p$ & $K_i$ & $K_d$ \\
        \hline
        \rule{0pt}{2ex}Ziegler Nichols \cite{8} & $3.872$ & $8.031$ & $0.466$ \\
        \rule{0pt}{2ex}BFOA \cite{8} & $3.185$ & $4.672$ & $0.655$ \\
        \rule{0pt}{2ex}IMC \cite{5} & $0.666$ & $1.018$ & $0.223$ \\
        \rule{0pt}{2ex}IPSO \cite{10} & $3.935$ & $8.147$ & $1.576$ \\
        \rule{0pt}{2ex}MABC & $ 0.486$ & $1$ & $0.154$ \\
       \rule{0pt}{2ex}IPSO-DOBC & $3.935$ & $8.147$ & $1.576$ \\
        \hline
        \hline
    \end{tabular}
\end{table}
\begin{table}[H]
\caption{Parameters of PID Controller for AVR}\label{AVR_PID}\centering
    \begin{tabular}{c c c c}
        \hline
        \hline
         & $K_p$ & $K_i$ & $K_d$ \\
        \hline
        \rule{0pt}{2ex}Ziegler Nichols \cite{8} & $1.021$ & $1.874$ & $0.139$ \\
        \rule{0pt}{2ex}ABC \cite{9} & $1.652$ & $0.408$ & $0.365$ \\
        \rule{0pt}{2ex}PSO \cite{9} & $1.777$ & $0.382$ & $0.318$ \\
        \rule{0pt}{2ex}DEA \cite{9} & $1.949$ & $0.443$ & $0.342$ \\
        \rule{0pt}{2ex}NLTA \cite{6} & $1.299$ & $1.379$ & $0.788$ \\
        \rule{0pt}{2ex}BFOA \cite{8} & $0.788$ & $0.608$ & $0.335$ \\
       \rule{0pt}{2ex}NLTA-DOBC & $1.299$ & $1.379$ & $0.788$ \\
        \hline
        \hline
    \end{tabular}
\end{table}

\end{document}